# How the viscous subrange determines inertial range properties in turbulence shell models

Norbert Schörghofer, Leo Kadanoff, and Detlef Lohse *

June 28, 1995

The James Franck Institute, The University of Chicago,
5640 South Ellis Avenue, Chicago, IL 60637, USA

We calculate static solutions of the 'GOY' shell model of turbulence and do a linear stability analysis. The asymptotic limit of large Reynolds numbers is analyzed. A phase diagram is presented which shows the range of stability of the static solution. We see an unexpected oscillatory dependence of the stability range upon $\lg \nu$, where $\nu$ is the viscosity. This effect depends upon the discrete structure of the shell model and goes to zero as the separation between the shells is brought to zero. These findings show how viscous effects play a role in determining inertial properties of shell models and give some hints for understanding the effects of viscous dissipation upon real turbulence.

# 1 Introduction

Kolmogorov's classic 1941 paper [1] asserts that in the limit of large Reynolds number, the inertial range behavior is independent of the form and magnitude of the dissipation. Many approaches such as large eddy simulations (LES, [2]) or simulations with hyperviscosity [3] have been based upon this idea that the viscous subrange (VSR) simply serves as a drain to remove the energy carried down through the inertial subrange (ISR). However, recently Leveque and She [4] have cast doubt upon the independence of the ISR on the VSR energy sink. They suggested that energy cascaded through the inertial range tended to be reflected at the dissipative scale. This reflected current could reappear in the integral range and thus could affect the details of integral range behavior. In particular, through this mechanism ISR scaling exponents could become Reynolds number dependent as also suggested by the log similarity model [5].

A variety of approximate models have been developed in recent years to test our ideas about turbulence. These models are shell models in which the velocities are put on a fractal series of shells in wave vector space. The different shells are characteristized by wave vectors of the form

$$k_n = k_0 \lambda^n \tag{1}$$

with $\lambda$ being the ratio between shells. One of these shell models is the so-called GOY model, introduced by Gledzer [6], Yamada, and Ohkitani [7], which involves a forcing at large scales, a cascade of energy through a set of shells in an inertial range, and a dissipative dominance of shells beyond the inertial range. (For some more recent studies see for example [8, 9].) The GOY model is perfectly set up to test the hypothesis that inertial range behavior is independent of what happens in the VSR. In contrast to full numerical simulations [10] and to reduced wave vector set approximations (REWA) [11], even an semi-analytic approach will be possible.

In this paper, we test Kolmogorov's basic hypothesis – the independence of ISR quantities on the VSR – by looking at the *static* properties of GOY model behavior. As first pointed out by Biferale et. al. [12] this model has a static solution for certain values of parameters. In this paper, we find the natural static solution (section 2), define the dependence of this solution upon the model parameters (section 3,4) and do linear stability analysis to determine the range of stability of the static solution (section 5).

The major conclusion of this paper is that Leveque and She [4] were right for the GOY model at least. The inertial range behavior of the static solution does depend upon the strength and structure of the dissipation. As we develop below, both the static solution and its range of stability are – for small values of the viscosity – periodic functions of the logarithm of viscosity, with the period being proportional to the logarithm of $\lambda$. The fundamental source of this periodic behavior is the meshing between the onset of dissipation and the shell structure. Thus, for example, the dissipation of energy has a different value when the dissipative term becomes first large at a given shell or if conversely it becomes large between shells. As the range of stability depends on the viscosity and as inertial range scaling corrections to K41 (commonly called $\delta\zeta_p = \zeta_p - p/3$ for the p-th moment of the velocity) set in smoothly at the borderline of stability [12], this suggest that also the scaling corrections $\delta\zeta_p$ depend on viscosity, maybe even in the fully chaotic state, as already discussed in [4, 9]. Since the shells are unphysical, this mechanism for viscosity dependence is also unphysical. In fact, this periodic behavior disappears when the shell ratio goes to unity. However, some roughly analogous mechanism for coupling between inertial range and dissipative range behavior might even exist in real turbulence and thus contradict some portion of the lore surrounding the K41 paper. This item will be discussed in our conclusions (section 6).

# 2 The GOY model and its static solution

The basic idea of the GOY model is to represent the wavevector space by N geometrically scaling wavevectors $k_n$, $n = 1, 2, \ldots, N$ as in equation (1). A complex number $U_n$ represents the typical velocity on scale $k_n^{-1}$. It is coupled to next and next nearest neighbors. Large scales are forced. Viscous damping becomes effective on small length scales. To be more precise, the GOY toy dynamics for the three-dimensional Navier-Stokes equations is given by

$$\frac{d}{dt}U_n = i\left[C_n(U)\right]^* - \nu k_n^p U_n + F\delta_{n,4}. \tag{2}$$

Here the cascade term, $C_n(U)$, is a bilinear expression in $U$ which links neighboring shells. In the GOY model it has the form:

$$C_n(U) = k_n U_{n+1}U_{n+2} - \epsilon k_{n-1}U_{n-1}U_{n+1} - (1-\epsilon)k_{n-2}U_{n-1}U_{n-2} \tag{3}$$

while the terms in $F$ and $\nu$ are respectively representations of the large-scale forcing and the small-scale viscous dissipation. The parameters in the model are the scaling factor $\lambda$ between the shells, the parameter $\epsilon$ determining the ratio among the three cascade terms, and the viscosity $\nu$, which can be considered as inverse Reynolds number. Most of the calculations in this paper are carried out for "normal" viscosity ($p = 2$). Sometimes we examine the effect of hyperviscosity by looking at $p > 2$. Normal viscosity is to be assumed unless stated otherwise. The forcing $F$ and the wavenumber $k_0$ determine length and time scales, we make the conventional choices $k_0 = \lambda^{-4}$ and $F = 5\cdot(1+i)\cdot 10^{-3}$. In the inviscous, unforced case, equation (2) conserves the total energy $\sum_n |U_n|^2/2$. A second conserved quantity is $\sum_n |U_n|^2(\epsilon-1)^{-n}$ which can be identified with helicity if $\lambda = 1/(1-\epsilon)$ [9]. The number of shells $N$ is chosen big enough so that the energy on the last shells is practically zero. Much work has been done in recent years to understand the *dynamics* of equation (2) [8, 9, 13].

For the static case the complex equation (2) can be set in real form. A simple change in the phase of the $U_n$ can eliminate all the $i$'s in the equation and leave the result real. There is a periodic behavior in $n$ built into the inertial range, in which the behavior on the different shells repeat with a period three [9]. The change in phase is done with the basic period three [13] in the form $U_n = u_n e^{i\phi_n}$, $F = fe^{i\Phi}$. By taking $\Phi = \pi/4$ we make $f$ real and equal to $5\cdot\sqrt{2}\cdot 10^{-3}$ in the standard case. From dynamical calculations we find that the static solution of the complex dynamical equation (2) picks the phases

$$\phi_n = \begin{cases} 9\pi/8 & \text{for } n = 3, 6, 9, \ldots \\ \pi/4 & \text{for } n = 1, 4, 7, \ldots \\ \pi/8 & \text{for } n = 2, 5, 8, \ldots \end{cases} \tag{4}$$

So we also choose this very phases and thus equation (2) is replaced by the real equation

$$\frac{d}{dt}u_n = -C_n(u) - \nu k_n^p u_n + f\delta_{n,4}. \tag{5}$$

with $C_n(u)$ defined in equation (3). Equation (5) is by the way Gledzer's original model for two-dimensional turbulence [6].

A phase choice like (4) can turn every static solution into real form. This is an exact result for nonvanishing viscosity [14]. In the examined cases, for the above chosen phases, all $u_n$ are positive, apart from $u_3$ which is negative and very small in modulus. ($u_3$ goes to zero with $\nu \to 0$.)

In the static case, the left hand side of eq. (5) is set to zero. The result is a fourth order *difference equation* which will then require four pieces of boundary data. Our solution is obtained by using the boundary conditions $u_0 = u_{-1} = u_{N+1} = u_{N+2} = 0$. Other boundary conditions are possible, e.g., one could think of fixing $u_0 = 1$ rather than forcing the system, but qualitatively they lead to the same conclusions as those boundary conditions chosen here.

In order to find the solutions $u_n$ of these $N$ real equations, we first get a rough solution by iterating forward the time dependent equations (2) for some small value of $\epsilon$. After a while, the velocities approach a reasonably constant set of values. This roughly static solution is then refined by Newton's method. After the exact static solution is obtained for some set of parameters, $(\nu, \lambda, \epsilon)$, then we continuously vary one or the other parameter(s) in small steps to obtain neighboring solutions. Once solutions are obtained over a reasonably large range, we can use the theory outlined below to extend the solution to smaller values of the viscosity and to arbitrarily large numbers of shells. Once there are enough shells, adding more just extends the VSR but does not change the values of the velocities in the other shells.

Since these static equations are non-linear and of high order, one expects many solutions. In this paper, we focus upon the particular solutions which evolve naturally after a transient time. They all have the same character as those in the base case of $\lambda$ equal to 2 and small values of $\epsilon$. These solutions exist over a wide range of parameters but do not exist everywhere. For example, figure 1 shows the region in $\lambda$ and $\epsilon$ for which one can find solutions to eq. (5) for the standard case $\nu = 10^{-7}$, $f = 5 \cdot \sqrt{2} \cdot 10^{-3}$. The number of shells $N$ is picked sufficiently large. This figure also shows the stability of our solution, as indicated by linear stability analysis. (See our work in section 5 below.) The lines on the figure define the places where the first stability eigenvalues pass from their stable to their unstable regions. Thus, the regions to left of these lines are the stability domains for the static solutions. The left solid line holds for the stability of eq. (2), the middle for that of eq. (5). In general, instabilities in the phase $\phi_n$ occur for smaller $\epsilon$ than for the modulus $u_n$, so the former is most frequently left to the latter.

Figure 1 is quite complex, describing many processes at once. We summarize this figure by describing what happens along the cut on the figure at $\lambda = 2$. For small $\epsilon$ there is a stable static solution of eq. (2). This solution continues to exist all the way up to $\epsilon_{snb} \approx 0.44990622582$ where there is a saddle node bifurcation [15]. At this bifurcation point one eigenvalue (in this case the largest) of the Jacobian of equation (5) becomes zero and a unique curve of fixed points passes through the bifurcation point which lies on the side $\epsilon < \epsilon_{snb}$ [15]. The static solution of (5) is stable over the entire range $]0, \epsilon_{snb}]$. Yet this is not the general case. For different $\lambda$ figure 1 shows that there are many cases where equation (5) becomes unstable as well. For $\lambda = 2$ equation (2) undergoes a Poincare-Andronov-Hopf (short: Hopf) bifurcation [15] at $\epsilon \approx 0.36987669663 \approx 0.3699$. This transition was analyzed by Biferale et al. [12]. The next most unstable mode turns positive at $\approx 0.39560317884 \approx 0.3956$. If one requires as did Gledzer [6] that the solution be real, then these instabilities cannot appear. Thus we see some of the major process which will affect the transition to chaos in the GOY model.

There is another striking feature of figure 1. It contains many wiggles. Even more pronounced wiggles are seen in $\epsilon_{stab}$ as a function of $\lg \nu$, see figure 12 below. Wiggles of this kind were first observed in a numerical study carried out by one of us [14]. One of the primary purposes of this paper is to explain these oscillations. As we shall see below they are a consequence of the shell structure [16].

Next we describe the qualitative nature of the solutions obtained for very high values of $N$ and small viscosity. We follow the classical analysis and divide our description into three parts:

a. A forcing range, $n = 1, 2, 3, 4$.

b. An inertial range. In this ISR, the viscous term is much smaller than any one of the different cascade terms. These cascade terms then cancel against one another.

c. A dissipative range. In the VSR, the viscous term cancels again the single largest cascade term–and the other two cascade terms are much smaller.

Start with the inertial range. Here one can obtain [9] a simple asymptotic solution for the product:

$$S_n = k_{n-1} u_{n-1} u_n u_{n+1} \tag{6}$$

for $n > 4$ of the form

$$S_n = A + B(\epsilon - 1)^n. \tag{7}$$

Here $A$ and $B$ determine the fluxes of the two conserved quantities. We look at the case in which the second term in equation (7) decays rapidly with n. In this case, the energy flux

$$J_n = k_n u_n u_{n+1} u_{n+2} + (1 - \epsilon) k_{n-1} u_{n-1} u_n u_{n+1} \tag{8}$$

gives the crucial transfer of energy (and information) from one region of $n$ to another, namely

$$J_n = (2-\epsilon)A. \tag{9}$$

The other flux [9]

$$L_n = (\epsilon-1)^{-n}(k_{n-1}u_{n-1}u_n u_{n+1} - k_n u_n u_{n+1}u_{n+2}) \tag{10}$$

is

$$L_n = (2-\epsilon)B \tag{11}$$

and carries no additional information, as, see below, equation (15) $J_n/L_n = A/B = (1-\epsilon)^4$ holds.

So far we have described two constants $A$ and $B$ of the four constants of integration which describe the solution in the ISR. The two additional arbitrary constants arise from the symmetry of the ISR equations under a period three multiplicative modification of $u_n$. The ISR solution is invariant under the multiplicative modification of solutions in the form

$$u_n \to \begin{cases} \alpha u_n & \text{for } n = 4, 7, 10, \ldots \\ \beta u_n & \text{for } n = 5, 8, 11, \ldots \\ \gamma u_n & \text{for } n = 6, 9, 12, \ldots \end{cases} \tag{12}$$

Since the modifications of equation (12) also change $A$ and $B$, they are not independent of the previously defined constants of integration. However if one demands that the product of the three changes be unity:

$$1 = \alpha\beta\gamma \tag{13}$$

then this freedom will not change the value of $A$ or $B$. By demanding the constraint of equation (13) we then have four independent constants of integration, which we can pick to be $A$ , $B$, $\beta$ , and $\gamma$.

Notice the nature of the information transfer through the inertial range. Our boundary conditions fix four parameters. The value of $B$ is of minor importance, as, when calculating $u_n$ as a function of the four constants $A$, $B$, $\beta$, and $\gamma$, it will not depend on $B$ for large $n$. However, both the value of $A$ – which determines the energy flux and equally the values of $\beta$ or $\gamma$ – which determine the ratios $u_{3k+2}/u_{3k}$ and $u_{3k+1}/u_{3k}$ – have an entirely non-decaying effect throughout the inertial range. Thus these three parameters transfer information without loss up and back over the whole ISR.

To set the four parameters, the system sets two conditions on each boundary of the ISR. To understand the integral scale, we note that $u_3$ is very small and in fact goes to zero with $\nu$ as the viscosity approaches zero, as mentioned above. We thus neglect $u_3$ in all the static equations with $n$ greater than 2. The resulting simplified set of equations is essentially the static version of equation (5)

$$0 = -C_n(u) - \nu k_n^p u_n + f\delta_{n,4} \tag{14}$$

with modified integral scale boundary conditions $u_3 = u_2 = 0$. These boundary conditions are similar to those suggested by Biferale et al. [12]. One of these boundary condition sets the ratio of $A$ and $B$. To make equation (7) consistent with $u_3 = 0$, i.e., $S_4 = 0$ from equation (6), we must take

$$B = A/(1-\epsilon)^4. \tag{15}$$

The other boundary condition for the inertial end is obtained from equation (14) applied to $n = 4$. Since $u_3$ is negligible, we see

$$f = k_4 u_5 u_6. \tag{16}$$

Notice that equation (16) uses the applied forcing to fix a boundary condition upon the inertial range but that it does not in itself fix the value of the energy flux. To know that flux, we would need to know $u_4 f$ – which is the energy flux. However, we shall not know $u_4$ until we have solved the entire problem including the boundary condition at the end of the VSR.

Deep in the VSR, we demand that the velocity decreases rapidly with shell number. In fact, as we shall develop just below, for very large $n$ we can obtain the very rapid decrease of the form:

$$\ln u_n = -const \times (k_n)^g \tag{17}$$

where $g$ is greater than one. Its actual value is of no importance for our argument. To see how this form arises, assume a decrease of this form. Then the $(1-\epsilon)$-term dominates the cascade terms and balances against the viscous terms. We then find, assuming that $(1-\epsilon)$ is positive:

$$\ln u_{n-2} + \ln u_{n-1} = \ln u_n + \ln(4\nu k_n/(1-\epsilon)). \tag{18}$$

The last term is asymptotically negligible in comparison to the others. Throw it away, and one finds a very simple solution to the resulting Fibonacci equation [17], namely

$$\ln u_n = -a\left(\frac{1+\sqrt{5}}{2}\right)^n + b\left(\frac{1-\sqrt{5}}{2}\right)^n \tag{19}$$

where $a$ is positive. Thus, the second order difference equation (18) has a two parameter family of solutions which die off rapidly with n. The two parameters $a$ and $b$ are set by fitting onto the inertial range solution. Note however, that we have gone from a situation in which we generally have four constants of integration to one in which we have but two in the far reaches of the VSR. Where have the other two gone? They were, in fact, set by the condition that $u_n$ must decrease rapidly with $n$. This condition eliminated two growing solutions. The demand that the inertial range solution must not excite the growing modes in the VSR then sets the two remaining parameters in the ISR solution. In this way, all parameters in the ISR solution are set. In particular, the energy flux is set by this interplay of ISR and VSR behavior.

## 3 Viscosity $\nu$ and shell distance $\lambda$ dependence of the solution

We now explore the $\nu$ and $\lambda$ dependence of the static solution. We saw in the last section that the solution was determined by detailed balancing of the various cascade terms and the dissipation term on the boundary between the VSR and ISR. The result (and hence the values of the flux) will depend in detail upon how the dissipation 'cuts in'. The detailed values of the integration constants will be different depending upon whether the dissipation first becomes important on a shell for which $n$ equals $3k$ or $3k+1$ or $3k+2$. The result will return to the same form only after the value of $\nu$ is decreased by a sufficient factor to bring the solution to the same form three shells down. Since the cascade term will increase by a factor $\lambda$ over three shells while the dissipation term will increase by a factor $\lambda^5$ over the same interval, then decreasing $\nu$ by a factor of $\lambda^{-4}$ will bring the system back to the same balance of terms three shells deeper down. To see this we plot in figure 2 the solution $u_n$ for a succession of cases in which $\nu$ differs by a factor of $\lambda^4$. The picture shows the same transition from ISR to VSR occurring again and again for the different values of $\nu$ – only for smaller $\nu$ the transition occurs on higher shells.

If we vary $\nu$ by some factor which is not an integer power of $\lambda^4$, the detailed structure of the wiggles in figure 2 will be changed because there is a change in the way the integral scale merges into the VSR. To see how this variation works, we look at the standard case $\lambda = 2$, $\epsilon = 0.3$ and vary the viscosity. In figure 3, we show the result of this variation upon the important determinants of ISR behavior: the energy current $J_n$ and the ratios $u_{3k+2}/u_{3k}$ and $u_{3k+1}/u_{3k}$. Each of these quantities, for $3k$ in the ISR, becomes independent of the subscript. Note the periodic oscillation of these quantities with $\lg \nu$. As we have already discussed the period is $\lg \lambda^4$. For the hyperviscous case $p = 4$, the same argument gives a period $\lg \lambda^{10}$; in general, it is $\lg \lambda^{3p-2}$.

Note the large relative amplitude of the oscillation. For figure 3 we have $\Delta J_n/J_n \approx 0.35$. To understand this result, we calculate the dependence of $J_n$ on $\lambda$ for fixed $\nu$, see figure 4. This oscillation comes from the dissipation of the energy current $J$ which takes place at the high-$n$ edge of the ISR within a fixed k-range $\Delta k$ of, say, about one decade. If there are enough shells in this range, i.e. if $\lambda$ is close to 1, then $J$ can precisely adjust to the supplied viscosity. If, on the other hand, the shells are sparse, the total dissipation and thus the total flux $J$ sensitively depends on the exact positioning of the shells which are situated *within* the fixed range $\Delta k$. We can analyze the amplitude of the oscillations when the number of shells in this range is not too small. Since the motion caused by a variation in $\nu$ can at most move one shell into or out of range, we may expect the relative variation in dissipation caused by the motion to be the inverse of the number of shells in the range. The number of shells $\Delta n$ within $\Delta k$ is $\Delta n = \lg(\Delta k)/\lg\lambda$. Therefore we expect

$$\frac{\Delta J}{J} \sim \frac{1}{\Delta n} = \frac{\lg\lambda}{\lg(\Delta k)} \approx \lg\lambda. \tag{20}$$

When we look at the numerics, we find order of magnitude agreement with equation (20). For $\lambda = 2$, we may expect an oscillation $\Delta J/J \approx \lg 2 \approx 0.3$, as observed. But we also observe some non-analyticity as $\lambda$ goes to 1. Such non-analyticities often arise in continuum limit problems, see e.g. ref. [18].

# 4 Large Forcing solution

So far all of our analysis has been performed for large-$N$ systems. As $\lambda$ got close to one we increased the number of shells so that, in effect, we were always working in the limit of infinite shell numbers. However, the matching between shells and dissipation which we are discussing in this paper is very well illustrated by considering the case of the shell number $N$ fixed. If we then make the forcing constant $f$ very "large", nearly all of the dissipation will occur on the last two shells. This gives us the definition of the critical $f = f_c$, beyond which we consider $f$ as large. We define $f_c$ by the condition

$$\nu k_N^2 u_N^2 \sim f_c u_4. \tag{21}$$

With Kolmogorov scaling $u_N \sim u_4(k_N/k_4)^{-1/3}$ and $u_4 \sim \sqrt{f/k_4}$ on dimensional grounds we obtain

$$f_c = k_0^3 \nu^2 \lambda^{(8N+4)/3} \tag{22}$$

as critical value for $f$.

In figure 5 we display the $f$ dependence of $u_4$, $u_5$, and $u_6$ for the standard case $N = 22$, $\lambda = 2$, $\epsilon = 0.3$, $\nu = 10^{-7}$. Since the coefficients $A$ and $B$ have the same $f$-dependence, we can restrict ourselves to theses three $u_n$; all higher $u_n$ of the same triality have the same scaling dependence as the corresponding basic velocity $u_4$, $u_5$, or $u_6$. We observe two quite different scaling regimes in figure 5. Indeed, the crossover between them agrees with what we calculate from equation (22), namely $f_c = 2.8$.

For small $f < f_c$, all $u_n$ ($n \neq 3$) scale as $u_n \sim f^{1/2}$, apart from wiggles of period $\lambda^8$ in $\lg f$. The $u_n \sim f^{1/2}$ scaling follows from dimensional analysis. Lengths are measured in terms of $[k_0^{-1}]$, times in terms of $[fk_0]^{-1/2}$. So velocities are measured in terms of $[k_0^{-1/2} f^{1/2}]$ and thus scale as $u_n \sim f^{1/2}$. From equation (5) it follows that $u_3 = \nu k_1 u_1/u_2 \sim f^0 \sim 1$. The wiggles on the underlying scaling law mirror the above explained fluctuations in $\nu$ for fixed $f$. As $[\nu] = \mathit{length}^2/\mathit{time} = [k_0^{-3/2} f^{1/2}]$, the period $\lambda^4$ in $\lg\nu$ means a period $\lambda^8$ in $\lg f$ which exactly is what we observe.

For large $f > f_c$ the dissipative mechanism is quite different from the Navier-Stokes case, yet it is similar to large eddy simulation-type models (see e.g. [2]) with some energy drain for large wavevectors. We start from K41 scaling, which can be expressed as

$$u_n = \lambda u_{n+3}. \tag{23}$$

Thus, we focus upon predicting the $f$-scaling of $u_{3k+1}, u_{3k+2}, u_{3k}$. Surprisingly, it depends on the total number $N$ of shells. The reason is that due to the $n \to n+3$ symmetry it matters whether exclusively complete triples $(u_n, u_{n+1}, u_{n+2})$ fit between the forced shell $n = 4$ and the dissipative shells $n = N, N-1$, or whether the last triple has to be truncated to $(u_n, u_{n+1})$ or even to $(u_n)$. Note that the *total* number of complete triples between $n = 4$ and the dissipative shells $N, N-1$ does *not* matter. We thus have to distinguish between the cases $N = 3k+1$, $N = 3k+2$, and $N = 3k$. The constraints on the $u_n$'s are the same in all three cases. Equation (16) provides a small-$n$ constraint. The large $n$-relations are obtained from equation (14) applied for $n = N$ and $n = N-1$. The resulting constraints are

$$u_N \sim u_{N-1} u_{N-2}, \tag{24}$$

$$u_{N-1} \sim u_N u_{N-2}, \tag{25}$$

but via the $n \rightarrow n+3$ symmetry they translate to quite *different* $f$-scaling for $u_4$, $u_5$, and $u_6$ for the cases $N = 3k+1, 3k+2, 3k$, which are displayed in table 1.

The spectrum for the case $N = 3k+1 = 22$ is shown in figure 6. It again reveals the period 3 structure of the GOY equations. As follows from table 1, the amplitude of this oscillation linearly grows with $f$.

The analysis of this chapter again reveals how the detailed structure of the VSR and the stirring subrange SSR determine the structure of the whole solution throughout. One should not be too surprised: Since equation (5), as discussed above, can be considered as a difference equation in the variable $n$. Again its solution is determined by the left and right boundary conditions for the equations for $n = 1, 2$ and $n = N, N-1$.

# 5 Stability analysis

Up to now we only discussed the existence of the solutions and its dependences on various parameters, but not whether the solution is linearly stable. Yet the case treated in recent publications [8, 9] is the unstable one, since it is only in this region that the GOY equations make sense as a model for turbulence. We would like to explore how unstability is achieved and how it depends on the various parameters. The first stability analysis for the GOY systems was performed by Biferale et al. [12] for one set of parameters $N = 19$, $\lambda = 2$, $\nu = 10^{-6}$, $k_0 = 0.05$. They find that instability in the GOY equations (5) sets in via a Hopf bifurcation (i.e., a complex conjugate pair of eigenvalues passes through the imaginary axis) at $\epsilon = \epsilon_{stab} \approx 0.38$. We find that this mechanism is rather general.

To discuss the stability of the GOY equations, we return to the consideration of equation (2). We are interested in the stability properties of the complex solution $U_n = u_n \exp(i\phi_n)$. Variations in $U_n$ are split in those in the modulus

$$u_n = u_n^{(0)} + \delta u_n \tag{26}$$

and those in the phase

$$\phi_n = \phi_n^{(0)} + \delta \phi_n. \tag{27}$$

The stationary phase $\phi_n^{(0)}$ is given by equation (4). The stability matrix equation for the modulus is readily obtained from equation (5) as

$$\frac{d}{dt}\delta u_n = -\Sigma(D_{nm} + C_{nm})\delta u_m. \tag{28}$$

Here $D_{nm}$ and $C_{nm}$ are respectively the dissipation and the cascade contributions to the linear response defined as

$$D_{nm} = \nu \delta_{nm} {k_n}^p \tag{29}$$

and

$$C_{nm} = \frac{\partial}{\partial u_m} C_n(u). \tag{30}$$

To obtain the stability matrix equation for the phase we first derive the phase dynamical equation which corresponds to equation (5) for the modulus and which determines the phase variation $\delta\phi_n$ around the stationary solution $(u_n^0, \phi_n^0)$,

$$\begin{aligned} u_n \exp(i\delta\phi_n)\frac{d}{dt}(\delta\phi_n) &= -\nu k_n^p \exp(i\delta\phi_n) + f\delta_{n,4} \\ &- k_n u_{n+1}u_{n+2}\exp(-i(\delta\phi_{n+1}+\delta\phi_{n+2})) \\ &+ \epsilon k_{n-1}u_{n-1}u_{n+1}\exp(-i(\delta\phi_{n-1}+\delta\phi_{n+1})) \\ &+ (1-\epsilon)k_{n-2}u_{n-1}u_{n-2}\exp(-i(\delta\phi_{n-1}+\delta\phi_{n-2})). \end{aligned} \tag{31}$$

Here we have dropped the index zero at the stationary solution $u_n^{(0)}$ for convenience. Linearization of (31) gives

$$\frac{d}{dt}\delta\phi_n = -\Sigma(D_{nm} - C_{nm}^{\phi})\delta\phi_m \tag{32}$$

with

$$C_{nm}^{\phi} = \frac{u_m}{u_n}\frac{\partial}{\partial u_m}C_n(u). \tag{33}$$

A similarity transformation of $D_{nm} - C_{nm}^{\phi}$ with the matrix $\delta_{nm}u_m$ leaves the eigenvalues unchanged and simplifies equation (32) to

$$\frac{d}{dt}\delta\phi_n = -\Sigma(D_{nm} - C_{nm})\delta\phi_m, \tag{34}$$

which only differs by a sign from equation (28).

Both of the stability matrices $-D_{nm} - C_{nm}$ and $-D_{nm} + C_{nm}$ have their own set of eigenvalues and eigenstates. As usual the eigenvalues, called $\sigma$, may be real or complex. The complex eigenvalues do not imply anything about the complexity of the perturbations around the static-velocity, which must be real. Rather a pair of complex conjugate eigenvalues represents a pair of perturbations which convert up and back into one another as time goes on and thereby produce an oscillatory response.

To get numerical results, the eigenvalues of the non Hermitian Jacobians are calculated with the EISPACK package which is available via netlib. To achieve sufficient numerical accuracy for the eigenvectors quadruple precision turns out to be necessary.

When the real part of the eigenvalue is smaller than zero, the perturbation produced by the corresponding eigenstate decays. Conversely, when the real part is positive, the perturbation produces an exponentially growing response. For this reason, we focus our attention upon the eigenvalues with the largest real part.

There are a large number, $2 \cdot N$, of eigenvalues. Figure 7 shows the pattern of eigenvalues for a typical situation ($\lambda = 2$, $N = 104$, $\epsilon = 0.3$, $\nu = 10^{-7} \cdot 16^{-19} = 1.32349 \cdot 10^{-30}$). Notice how the eigenvalues arrange themselves into families. Each right eigenstate has a peak at some value of $n$ (Fig. 8b). Call this value $m$. Since the eigenvalues represent a relaxation rate, it is not surprising that for values of $m$ toward the middle of the inertial range both the eigenvalues of the two branches in figure 7 have the magnitude $\sigma \sim {k_m}^{2/3}$ as predicted by K41 scaling. The real family of eigenvalues continues into the dissipation range where $\sigma \sim {k_m}^2$ since this is the magnitude of the dissipative relaxation rate. Only a few of the eigenvalues fail to fall into the pattern described here. These special eigenvalues have right eigenstates which tend to have a large weight at low shells.

One phase disturbance has eigenvalue zero. This is reflective of a remaining phase ambiguity [13, 14] in the solution. Starting from any solution of our static equations one can obtain another one by the following process: All the $u's$ with the same triality as $u_5$ are modified by multiplication by $e^{i\phi}$ (with $\phi$ real) while all the ones with the triality of $u_6$ are given the opposite phase. The new set of $u_n$ is also a solution. This ambiguity in the solution is reflected in a zero eigenvalue, which will have no further importance for us.

For most values of $\lambda$ (the only exception being around $\lambda = 2.8$) the largest real part occurs in the phase perturbation so that the phases show a greater tendency toward instability. As a result, the stability bounds shown in figure 1, which describe where the first eigenvalues go unstable, are phase instabilities. Each of these is also a pair of complex conjugate eigenvalues. Thus, the leading instabilities of the static solution are actually Hopf bifurcations in the phase variables. One such bifurcation gives periodic oscillatory behavior. According to the Ruelle Takens analysis, two such Hopf bifurcations can give chaotic behavior. Evidently, the Hopf bifurcations can lead to the GOY-model chaos. Indeed, this scenario has been demonstrated by Biferale et al. [12] for the GOY model.

It is remarkable that the eigenvalue spectrums of the phase stability matrix $C_{nm} - D_{nm}$ and of the (negative) magnitude stability matrix $C_{nm} + D_{nm}$ are so different as the two matrices only differ by the dissipation contribution (29). This again is an indication for our main point of this paper: Dissipation effects can considerable change dynamical properties even in the inertial range.

Luca Biferale [19] predicted that the GOY dynamics should turn unstable when the magnitude of the summands $|U_n|^2(\epsilon - 1)^{-n}$ in the second conserved (in the inviscous, unforced limit) quantity [9] $\sum_n |U_n|^2(\epsilon - 1)^{-n}$ *increases* with $n$. Plugging in K41 scaling, one obtains that the borderline of stability is

$$\lambda = (1 - \epsilon)^{-3/2}. \tag{35}$$

We test this prediction by plotting the line of first and second phase instability in a log-log plot of $\lambda$ against $1 - \epsilon$, cf. figure 9. Equation (35) catches the rough shape of the instability curves. This log-log plot also reveals that the stability curve *crosses* the curve of conserved helicity. It seems to oscillate around it and might asymptotically approach it.

The largest real parts of the eigenvalues of the phase Jacobian are shown in figure 10, again for the standard parameters $\nu = 10^{-7}$, $\lambda = 2$, $N = 22$, but any larger $N$ will lead to the same result, see above. In figure 11 we show how the eigenvalues $\sigma$ for the phase Jacobian matrix move within the complex $\sigma$-plane for increasing $\epsilon$. The above mentioned Hopf bifurcation is clearly seen. As mentioned in section 2, the solution ceases to exist around $\epsilon_{snb} \approx 0.4499$. For the $\epsilon$-range $]0, 0.4499]$ shown in figure 10, the modulus equations are linearly stable for these parameters. Yet this is not generally true: e.g. for $\nu = 10^{-7}$, $\lambda = 1.13$, $N = 104$ (or larger), instability occurs around $\epsilon^{\phi}_{stab} \approx 0.11$ in the phase equations and around $\epsilon^{u}_{stab} \approx 0.29$ in the modulus equations.

In figure 12 we present $\epsilon_{stab} = \epsilon^{\phi}_{stab}$ as a function of $\lg \nu$ for $\lambda = 2$. Strong oscillations occur which mirror the oscillations of the $u_n$ with $\lg \nu$ discussed in section 3. Therefore, we expect a period of $\lambda^4$, which indeed is the case. Our finding also suggests that the inertial range scaling exponents $\zeta_p$ of the p-th velocity moments depend on the viscosity $\nu$, as deviations $\delta\zeta_p = \zeta_p - p/3$ from the K41 value $\zeta_p = p/3$ set in smoothly at $\epsilon^{\phi}_{stab}(\nu)$. Such a dependence was already discussed in [4, 9].

The small wiggles on the curve in figure 12 are real. We only understand them in part. That every second oscillation should look the same can be understood along Biferale's above mentioned argument [19] for the onset of instability: It will make a difference whether viscosity cuts in a shell with positive $|U_n|^2(\epsilon - 1)^{-n}$ (the summands of the second conserved quantity) or negative one, i.e., odd or even $n$.

In figure 1 we have already presented $\epsilon^{\phi}_{stab}$ and $\epsilon^{u}_{stab}$ as a function of $\lambda$ for $\nu = 10^{-7}$. If turned by 90 degrees, this figure resembles figure 4, the increasing fluctuations of $u_n$ and $J_n$ with increasing $\lambda$.

For larger $\lambda$ the fluctuations of $u_n$ become so large that stability and unstability change more than once with increasing $\epsilon$ [14].

# 6 Summary and conclusions

We set out in this paper to understand qualitatively the complex phase diagram of figure 1. We achieved that goal by explaining the observed oscillations in that figure in terms of a particular and peculiar form of interaction between the integral scale and the VSR.

Thus, in this paper, we have discussed a special coupling among the different turbulent scales. In this form of coupling, the integral scale produces an energy flux but does not fully determine its value. Instead the flux-value is set by the joint effect of the production and the dissipation mechanisms, i.e., though *describing* the ISR, it is *set in* the VSR. In the present calculations for the GOY model, the ISR-VSR coupling was made manifest by oscillations of the flux (and maybe also of inertial range scaling exponents $\zeta_p$) as a function of viscosity. These oscillations in turn result from dividing the wavevector space in finite shells and vanish only for diminishing shell distance. In shell models the oscillations are unavoidable, though we expect them to be much less pronounced (if noticeable at all) in the REWA calculations [11] as many more modes and couplings are present. In real turbulence there are no shells and no such oscillations. Nonetheless the more general lesson of this calculation might apply and there might well be a similar cooperative behavior between the VSR and the ISR, since both the corrections to scaling (if they exist) and the multifractal components (if they exist) are similarly undetermined by the inertial range as is the energy flux and the scaling corrections in the GOY model. All these quantities are set by the action of both the integral scale and the dissipative scale – working together. An independent hint for such a mechanism is experimentally found by Castaing and collaborators [5] who clearly detect a Reynolds number dependence of ISR scaling exponents.

Thus turbulent properties might well be a product of the detailed way the viscous range fits onto the ISR and turbulence might in end have a richer dependence upon viscous properties than contemplated in K41. If this finding can be confirmed, it will have prime importance for all models and simulations where the small scales are not fully resolved, as for example in hyperviscosity calculations or large eddy simulations.

**Acknowledgements:** This work has been done with the support of the ONR. One of us (NS) was supported in part by the Austrian Federal Ministry for Science and Research. We thank L. Biferale, H. Grosse, I. Procaccia, Z.S. She, K. R. Sreenivasan, J. Wang, and Siegfried Grossmann for helpful discussions. It is our additional pleasure to congratulate Professor Grossmann on the occasion of his 65th birthday. We thank him for all that we have learned from him about science and turbulence, and for the excellent leadership he has shown in our field of activity.

## Tables

| | $N = 3k + 2$ | $N = 3k + 1$ | $N = 3k$ |
|---|---|---|---|
| $u_4$ | $\sim f$ | $\sim f$ | $\sim 1$ |
| $u_5$ | $\sim 1$ | $\sim f$ | $\sim \sqrt{f}$ |
| $u_6$ | $\sim f$ | $\sim 1$ | $\sim \sqrt{f}$ |
| $\epsilon_{diss}$ | $\sim f^2$ | $\sim f^2$ | $\sim f$ |
| $u_1$ | $\sim f^{-1}$ | $\sim f$ | – |
| $u_2$ | $\sim 1$ | $\sim f$ | – |
| $u_3$ | $\sim f^{-1}$ | $\sim 1$ | – |
| existence | $[0, \infty[$ | $[0, \infty[$ | $[0, \approx 1000]$ |
| stability | $[0, \infty[$ | $[0, \approx 12]$ | $[0, \approx 800]$ |

**Table 1**

Scaling properties of the solutions for the three different cases $N = 3k+2$, $N = 3k+1$, $N = 3k$ for large $f$. The first three lines follow from equations (24) and (25) and (23), the next one denotes the scaling of the total energy dissipation $\epsilon_{diss}$. The scaling of $u_1$, $u_2$, and $u_3$ follows from the first three equations of (5). Note that in the last case $N = 3k$ no solution exists. Correspondingly, the solution for the whole system of the $N$ real equations for $u_n$ ceases to exist for sufficiently large $f$. The approximate domain of existence and stability of the solution is given in the last two lines.

* On leave of absence from Fachbereich Physik, Universität Marburg, Renthof 6, D-35032 Marburg.

## Figures

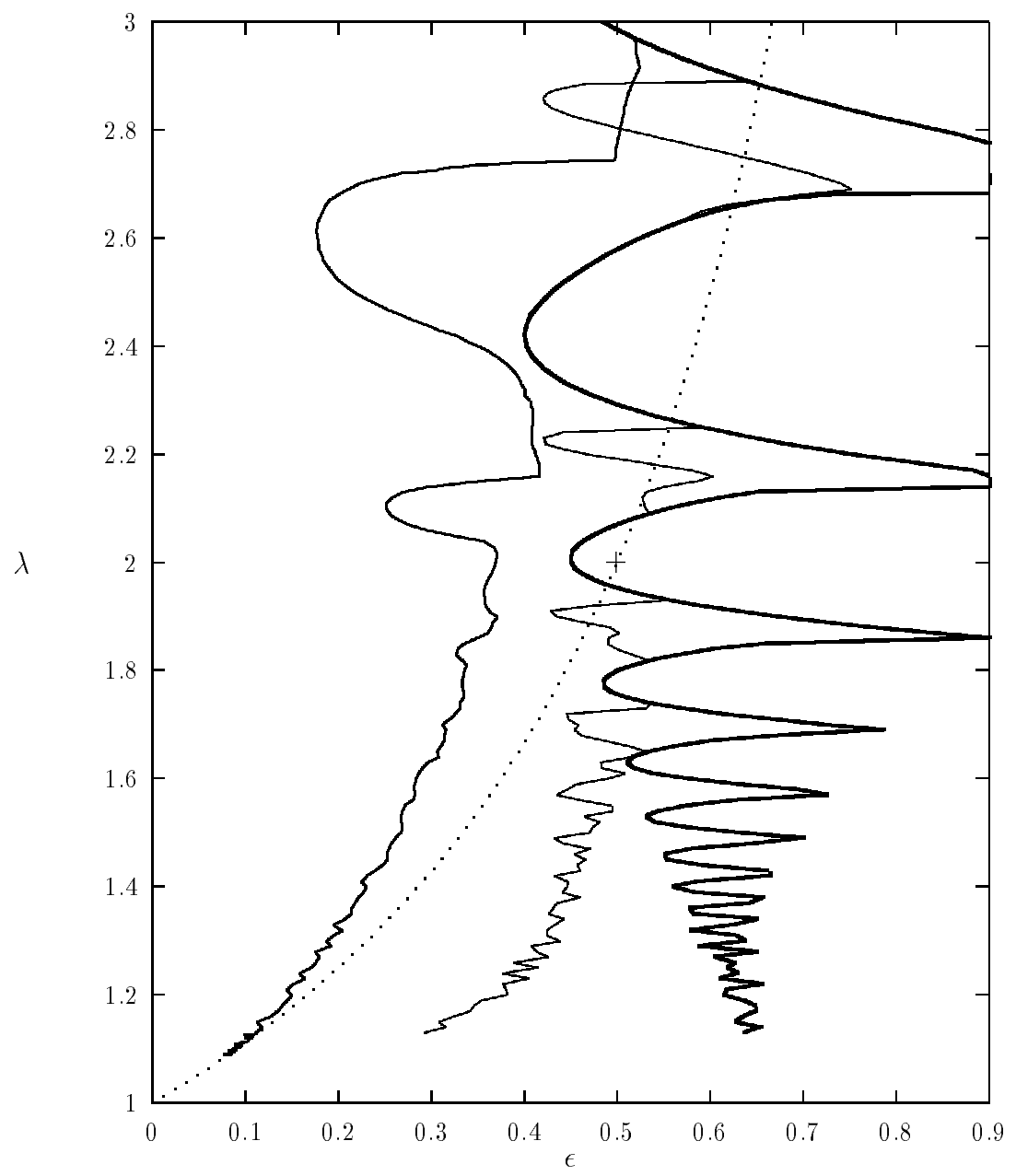

Figure 1: Phase diagram which shows regions of existence and stability of the static solution. The right solid line is $\epsilon_{snb}$, left of which the static solution exists. Note that the minima and maxima of $\epsilon_{snb}(\lambda)$ roughly agree with those of $J(\lambda)$, cf. figure 4. For completeness we mention that far right of this line there are (unmarked) regions where a stable solution of the GOY model again exists. The left solid line is $\epsilon^{\phi}_{stab}$ (first phase instability), the middle thin solid line is $\epsilon^{u}_{stab}$ (first magnitude instability). The calculation is done for $\nu = 10^{-7}$ and $N$ sufficiently large so that its value does not matter. The dotted line marks $\epsilon = 1 - 1/\lambda$ along which the helicity is conserved and along which dynamical scaling exponents $\zeta_q$ (beyond some threshold) are the same up to numerical accuracy [9]. The cross marks the standard case $\lambda = 2$, $\epsilon = 0.5$.

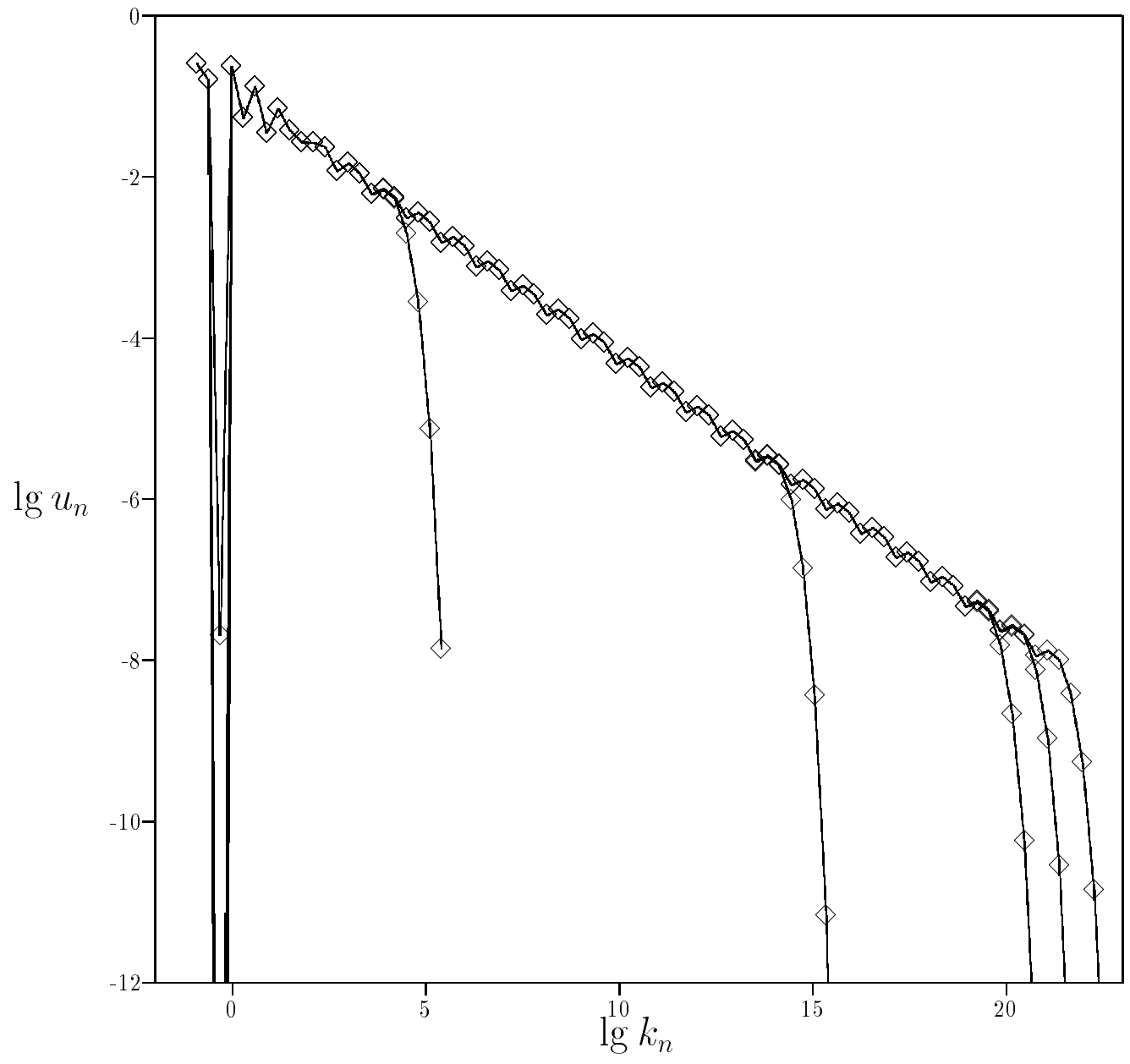

1

Figure 2: Spectrum $u_n$ vs $k_n$ for $\epsilon = 0.3$, $\lambda = 2$ for various different viscosities $\nu = \nu_0 = 10^{-7}$, $\nu = \nu_0 \cdot 16^{-11}$, $\nu = \nu_0 \cdot 16^{-17}$, $\nu = \nu_0 \cdot 16^{-18}$, $\nu = \nu_0 \cdot 16^{-19}$, left to right. Note that the detailed wiggle structure in the ISR, which now is identical for all cases, will be different when changing $\nu$ by a factor different from $(\lambda^4)^{-m}$ where $m$ is an integer.

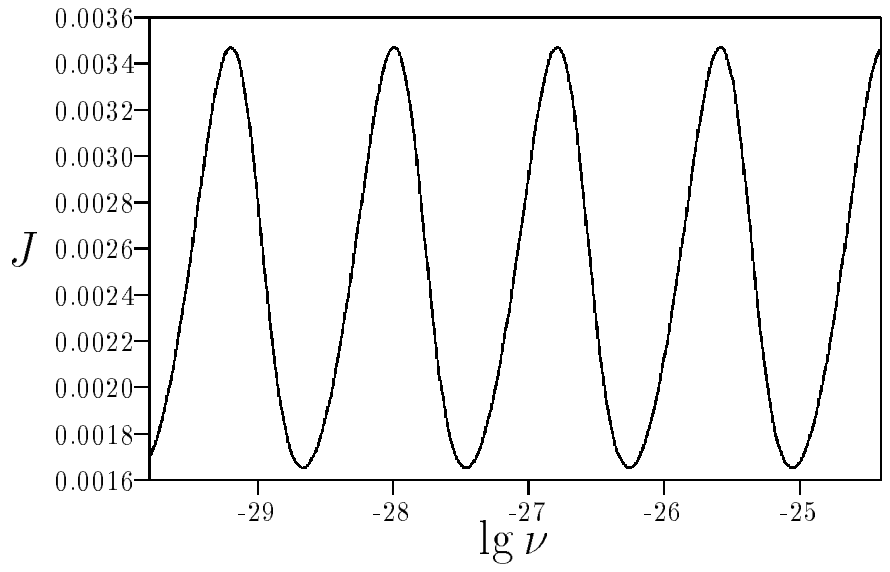

1

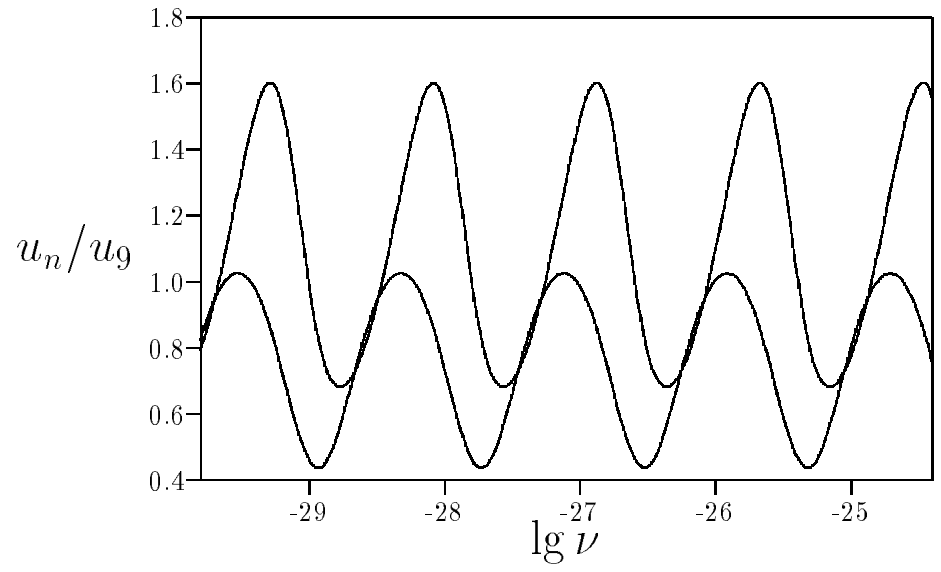

1

Figure 3: (a) The energy flux $J = J_n$ for $n$ in the ISR, showing strong oscillations with $\lg \nu$. The period is $\lambda^4$. Here we chose $N = 104$, $\lambda = 2$, $\epsilon = 0.3$. Note that the solution is unstable in some regions which we will discuss in detail in section 5. (b) The ratios $u_{3k+2}/u_{3k}$ (lower) and $u_{3k+1}/u_{3k}$ (upper) for $k = 3$, showing strong oscillations as $J_n$. For all $3k$ in the ISR, these ratios are the same.

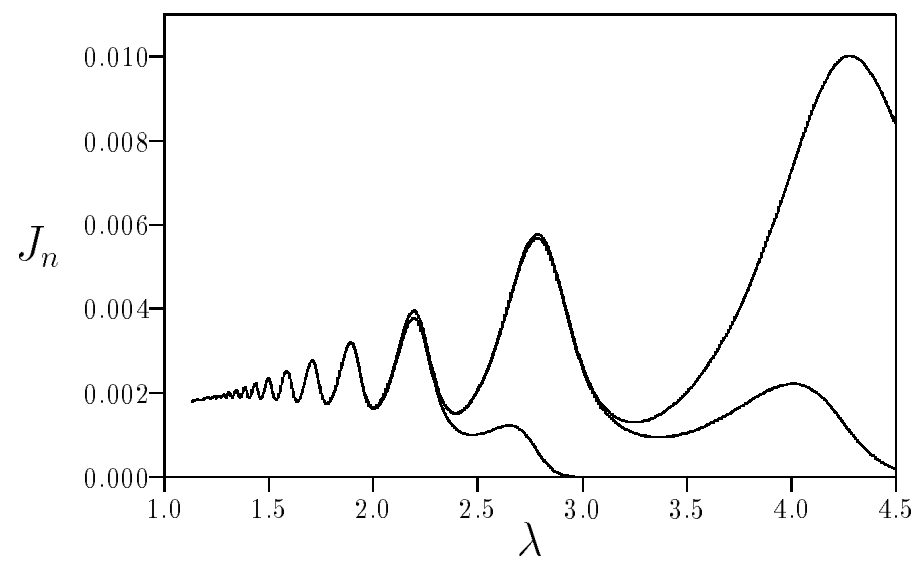

1

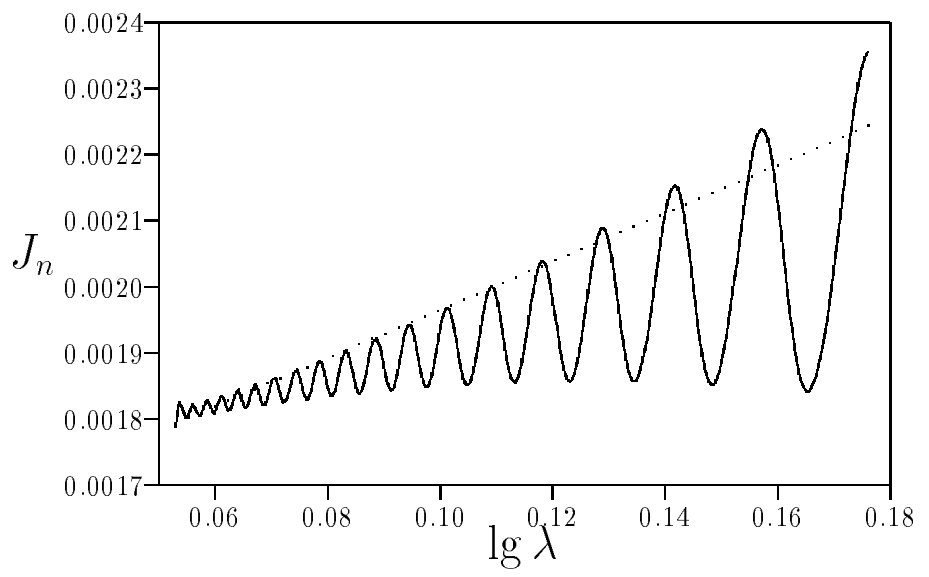

1

Figure 4: (a) The energy flux $J_n$ for $n = 4, 11, 14$, top to bottom. We chose $N = 104$, $\nu = 10^{-7}$, $\epsilon = 0.3$. The oscillations increase with increasing $\lambda$. For large $\lambda$ viscous effects can already be felt for the large shells $n = 11, 14$. (b) Enlargement of the left part of the figure 4a in linear-log scale. The dashed line touching the maxima is $0.0016 + 0.0004 \lg \lambda$, according to (20).

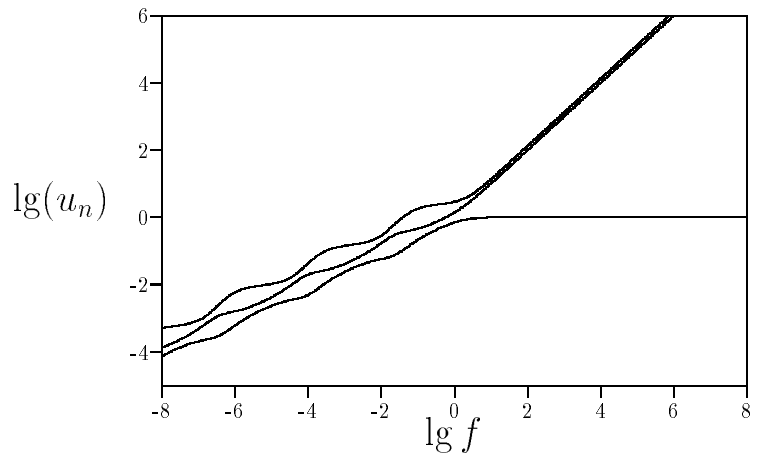

1

Figure 5: $u_n$ vs $f$ for $n = 4, 6, 5$ (top to bottom) for $N = 22$, $\epsilon = 0.3$, $\nu = 10^{-7}$. Two regimes are identified: A small $f$ regime with $\sqrt{f}$-scaling and wiggles of period $\lambda^8$, and a large $f$ regime with $f^1$ and $f^0$ scaling.

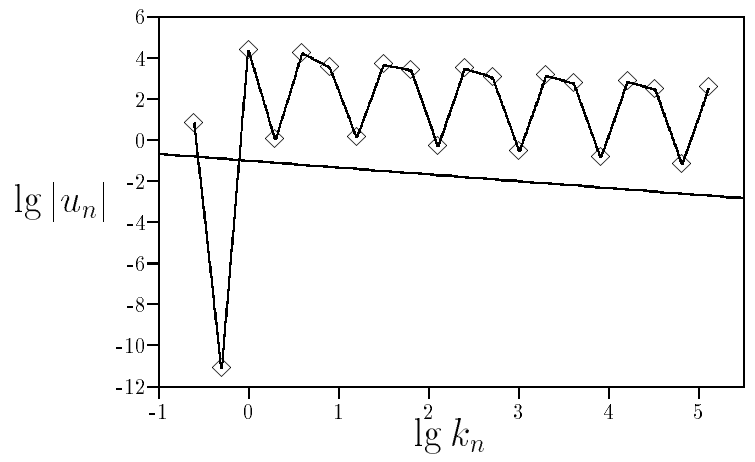

1

Figure 6: Energy spectrum $u_n$ vs $k_n$ in the large $f$ regime: $f \approx 10^4$, $\epsilon = 0.3$, $\nu = 10^{-7}$, $N = 22$, $\lambda = 2$. Also shown is the power law $k_n^{-1/3}$.

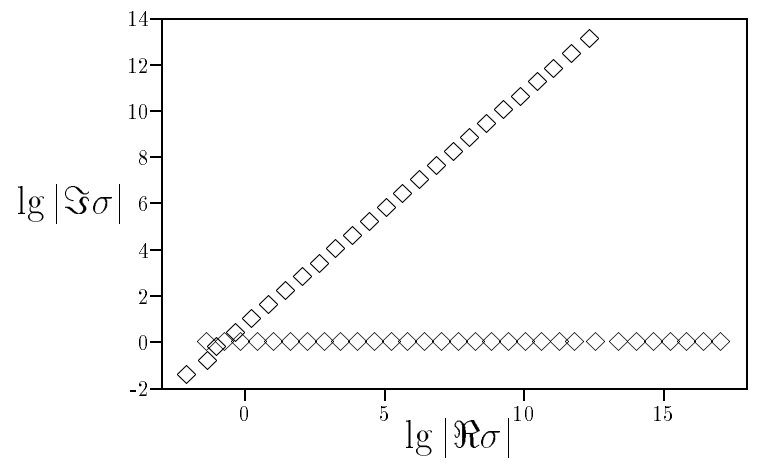

1

Figure 7: Eigenvalues for the matrix equation (28) of the modulus for $N = 104$, $\epsilon = 0.3$, $\lambda = 2$, $\nu = 1.32349 \cdot 10^{-30}$ in the complex plane. The complex pairs of eigenvalues $(\Re\sigma, \pm\Im\sigma)$ are visualized as $(\lg|\Re\sigma|, \lg(|\Im\sigma|)$, the real eigenvalues $\sigma$ as $(\lg|\sigma|, 0)$. The largest eigenvalue $\sigma \approx -10^{-6}$ which is associated with the irregularity at $u_3$ is not seen in this plot. It is this eigenvalue that turns zero at $\epsilon_{snb} \approx 0.4499$. Note that the irregularity in the pattern for $(\lg|\sigma|, 0)$ around 12 is not an numerical artefact but it is real and coincides with the end of the upper curve. Its physical meaning is the onset of the dissipation range VSR.

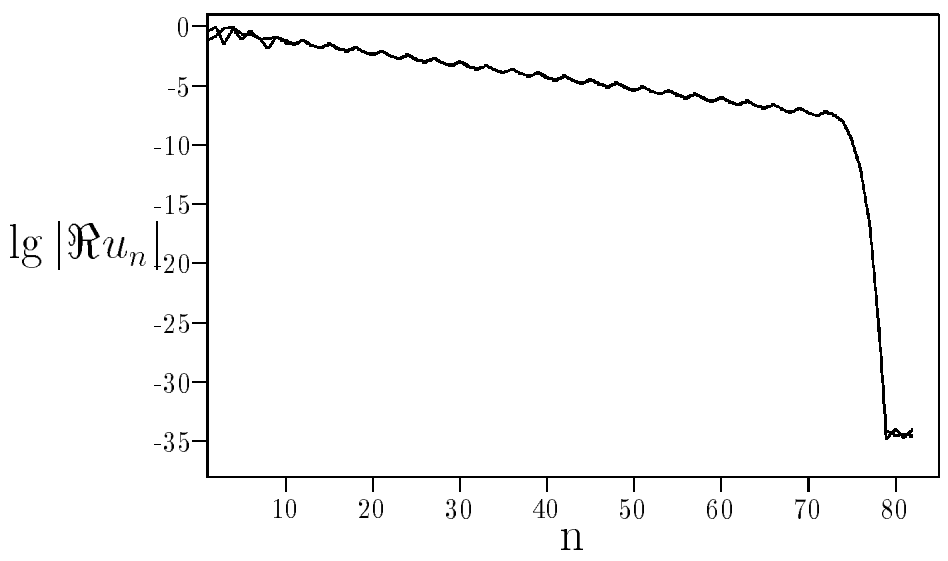

1

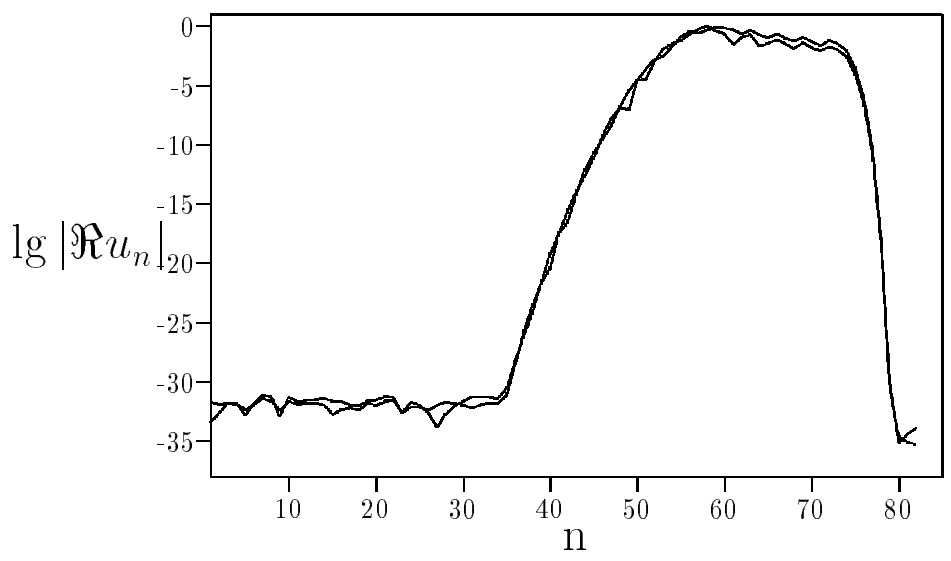

1

Figure 8: Eigenvector $|\Re(u_n)|$ vs. $k_n$ for magnitudes, $N = 104$, $\epsilon = 0.3$, $\lambda = 2$, $\nu = 1.32349 \cdot 10^{-30}$ for (a) the big eigenvalue $\sigma \approx -0.082 \pm i0.035$ (left and right eigenvectors) and (b) the smaller eigenvalue $\sigma \approx -2.0 \cdot 10^{9} \pm i1.0 \cdot 10^{10}$ (left and right eigenvectors). We did the calculation with quadrupole precision, the numerical noise level $10^{-32}$ can be recognized in both figures (a) and (b).

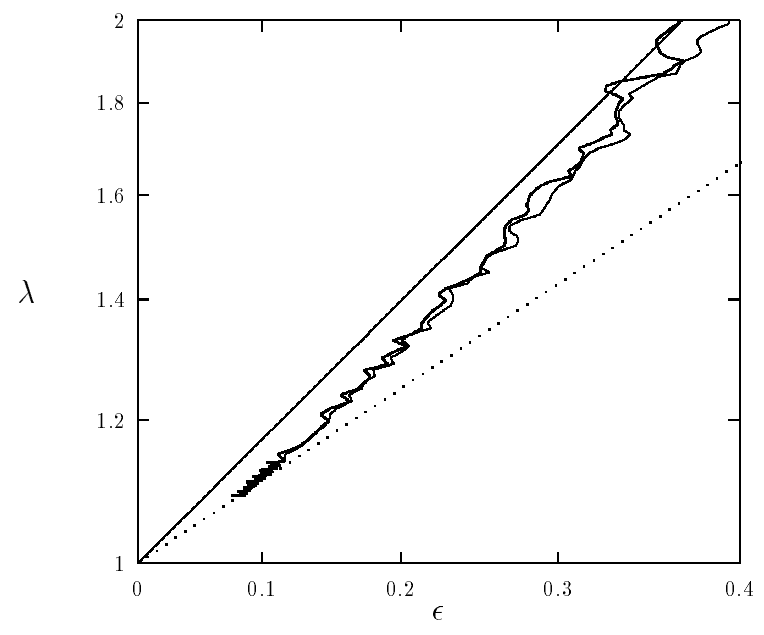

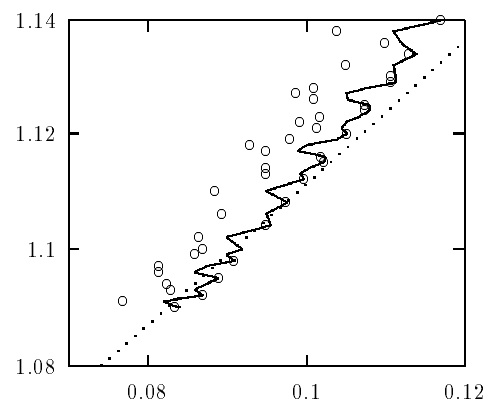

Figure 9: First and second phase (Hopf doublet) instabilities $\epsilon^{\phi}_{stab}$ in a lg-lg plot of $\lambda$ against $1-\epsilon$. Also shown is Biferale's prediction [19], equation (35) (solid), and the line $\lambda = 1/(1-\epsilon)$ along which "helicity" is the second conserved quantity [9] (dotted). The inset shows an enlargement of the small $\lambda$ range which shows that the instability line crosses the line of conserved "helicity". The calculated points for the first eigenvalue crossing zero are marked by dots; the line without dots denotes the second phase instability.

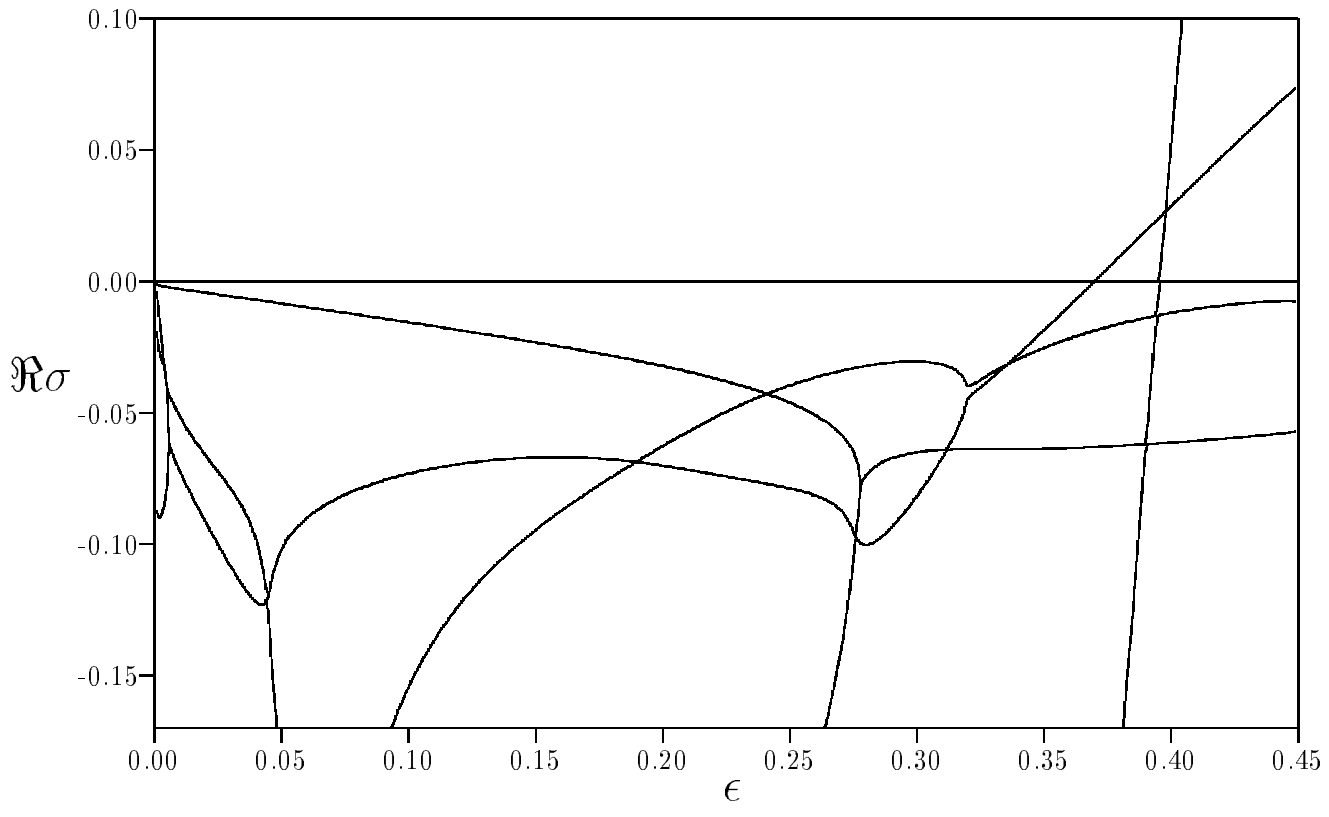

1

Figure 10: Largest eigenvalues of the phase Jacobian for the standard parameter case. The system turns unstable for $\epsilon \approx 0.3699$. A second mode becomes unstable at $\epsilon \approx 0.3956$. The capital letters refer to figure 11.

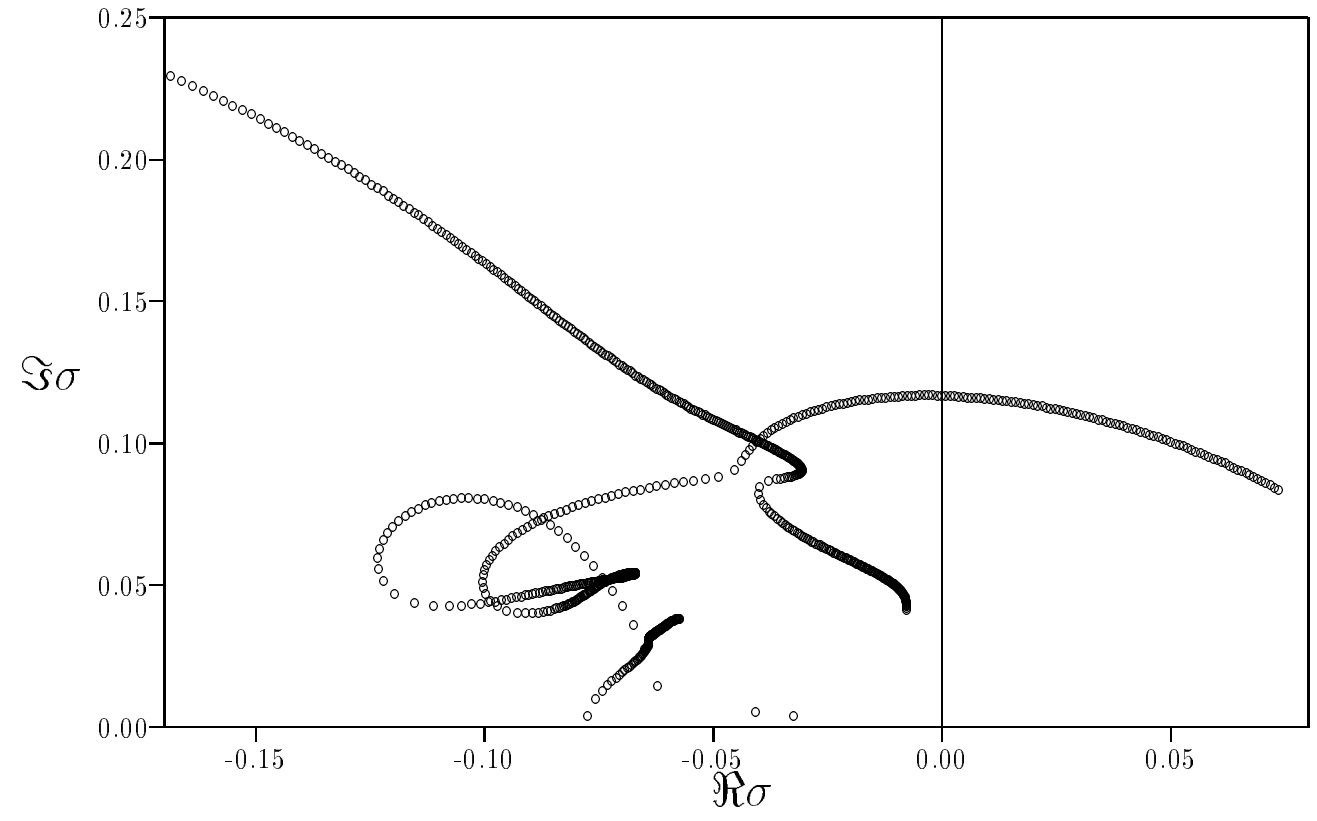

1

Figure 11: Movement of the eigenvalues $\sigma$ (of the phase Jacobian) in the complex plane for increasing $\epsilon$. We chose $N = 22$, $\nu = 10^{-7}$, $\lambda = 2$. The Hopf bifurcation at $\epsilon \approx 0.3699$ is clearly identified. The letters A, B, C denote the same eigenvalues as in figure 10. Eigenvalue D in that figure has much larger imaginary part $\Im\sigma \approx 2$ when crossing the imaginary axis and can thus also not be seen in this figure. Eigenvalue E in figure 10 (between zero and $\approx 0.23$) has zero imaginary part and can thus not be seen here. For this plot we chose $\Delta\epsilon = 0.01$. The lateral point density in each line of the plot is proportional to the inverse "speed" $d\Re\sigma/d\epsilon$ in figure 10. The four arrows correspond to $\epsilon = 0.1$, 0.2, 0.3, and 0.4.

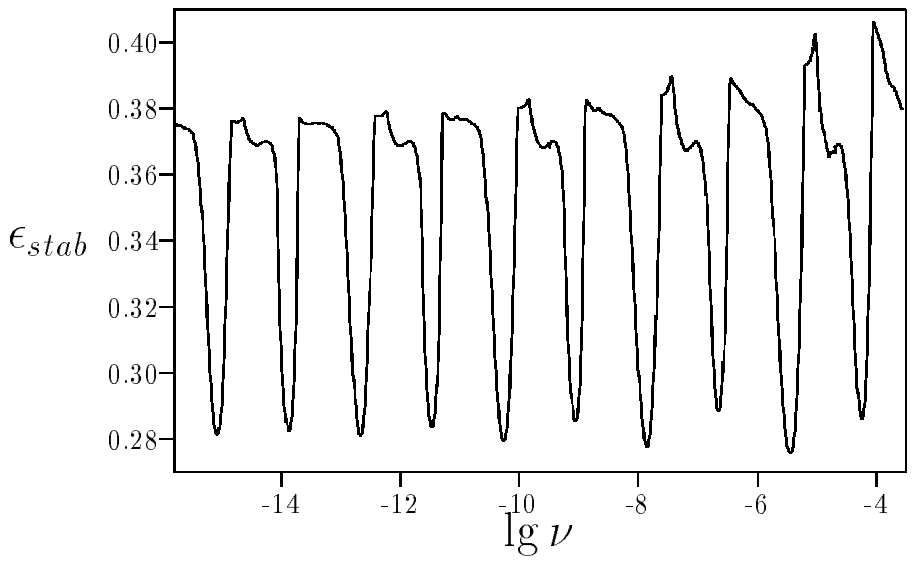

1

Figure 12: Dependence of $\epsilon_{stab} = \epsilon_{stab}^{\phi}$ on $\lg \nu$ for the standard parameter case $\lambda = 2$. The shell number $N$ is always chosen large enough so that we are in the infinite shell number case. Note that these fluctuations are much smaller for smaller $\lambda$, similar to those in the current $J_n$, cf. figures 3 and 4. They cannot account for the slight deviations between the Biferale prediction (35) and our numerical results in figure 9.